\documentclass[11pt]{article}
\usepackage{geometry,amsfonts,amssymb,amsmath,amsthm,cite}

\newcommand{\ft}[2]{{\textstyle\frac{{#1}}{{#2}}}}
\newcommand{\fft}[2]{{\frac{{#1}}{{#2}}}}
\newcommand{\tr}{\mbox{Tr\,}}

\newcommand\beq{\begin{equation}}
\newcommand\eeq{\end{equation}}
\newcommand\bea{\begin{eqnarray}}
\newcommand\eea{\end{eqnarray}}

\makeatletter
\@addtoreset{equation}{section}
\makeatother

\begin{document}
\setcounter{page}{0}
\begin{titlepage}
\titlepage

\begin{flushright}
IPhT-T10/161\qquad MCTP-10-50
\end{flushright}

\vspace{30pt}
\begin{center}

{\Large {\bf Computing $1/N^2$ corrections in AdS/CFT}}

\vspace{20pt}

James T. Liu$^a$ and Ruben Minasian$^b$

\vspace{20pt}

{${}^a$\it Michigan Center for Theoretical Physics\\
Randall Laboratory of Physics, The University of Michigan\\
Ann Arbor, MI 48109--1040, USA}

\vspace{10pt}

{${}^b$\it Institut de Physique Th\'eorique, CEA/Saclay\\
91191 Gif-sur-Yvette Cedex, France}

\vspace{40pt}

\underline{ABSTRACT}
\end{center}

Stringy corrections in AdS/CFT generally fall into the category of either
$\alpha'$ effects or string loop effects, corresponding to $1/\lambda$ and
$1/N$ corrections, respectively, in the dual field theory.  While
$\alpha'^3R^4$ corrections have been well studied, at least in the context
of $\mathcal N=4$ super-Yang-Mills, less is known about the $1/N^2$
corrections arising from closed string loops.  In this paper, we consider
AdS$_5\times\mathrm{SE}_5$ compactifications of the IIB string, and
compute the closed string loop correction to the anomaly coefficients
$a$ and $c$ in the dual field theory.  For $T^{1,1}$ reductions, we find
the string loop correction to yield $c-a=1/24$, which is the contribution to
$c-a$ of a free $\mathcal N=2$ hypermultiplet.  We also comment on
reductions to lower dimensional AdS theories as well as the nature of
T-duality with higher derivatives.

\vfill
\begin{flushleft}
{28 October 2010}\\
\vspace{.5cm}
\end{flushleft}
\end{titlepage}

\newpage

\section{Introduction}

While many important features of string theory may be investigated in its
low energy limit, it is often desirable to go beyond supergravity and to
examine distinguishing features that separate string theory from ordinary
supergravity.  Such stringy effects include Kaluza-Klein modes and
non-perturbative states such as D-branes as well as string worldsheet
effects arising from the string loop expansion and the $\alpha'$
expansion.  In fact, the latter $\alpha'$ expansion, which is equivalent
to a higher derivative expansion in the effective field theory, has attracted
much recent attention for multiple reasons.

{}From a quantum gravity point of view, higher derivative corrections
serves as a means of probing string theory at a fundamental level.  This
has been successfully applied to the study of stringy black holes and
higher derivative effects on black hole entropy \cite{Maldacena:1997de,Harvey:1998bx,Kraus:2005vz,Sen:2005wa}
(see {\it e.g.} \cite{Sen:2007qy,Castro:2008ne} and references therein).
Alternatively, higher derivative corrections also play an important role
in AdS/CFT, showing up as finite coupling and in some cases $1/N$ effects in
the dual field theory.

The string $\alpha'$ expansion naturally leads to a higher derivative
expansion in the effective field theory, with each factor of $\alpha'$
accompanied by two additional derivatives.  In most cases, the focus has
been on the gravitation action, including $\alpha'R^2$ terms in the
heterotic effective action and $\alpha'^3R^4$ terms in the type II theories.
However, the complete $\alpha'$ expansion involves not just Riemann terms
but all fields of the theory.  Such corrections in their entirety are rather
complicated, and often only incomplete information is known.  Nevertheless,
the first higher order curvature terms are generally well established, and
for non-flux backgrounds, they are often sufficient for most purposes.

The perturbative $\alpha'^3R^4$ corrections in type II string theory arise at
both tree level and one-loop order.  In the IIA case, the corrections take the
schematic form
\begin{equation}
S_{\mathrm{IIA}}[\alpha'^3]=\fft{\alpha'^3}{2\kappa_{10}^2}\int d^{10}x
\sqrt{-g}\left[e^{-2\phi}(t_8t_8+\ft18\epsilon_{10}\epsilon_{10})R^4
+(t_8t_8-\ft18\epsilon_{10}\epsilon_{10})R^4+B\wedge t_8R^4\right].
\label{eq:apm3}
\end{equation}
The $t_8t_8R^4$ terms were first obtained by direct calculation of
four graviton scattering at tree level \cite{Gross:1986iv} and one loop
\cite{Sakai:1986bi} order, while the $\epsilon_{10}\epsilon_{10}R^4$
terms are related to the eight-dimensional Euler density, and first arise
at the level of five graviton scattering.  The one-loop CP-odd term 
  $B\wedge X_8$ is related to the five-brane anomaly, and
computed in \cite{Vafa:1995fj,Duff:1995wd}.  This is the IIA analog of the heterotic
Green-Schwarz term, and can be computed directly through a parity violating
five-point amplitude \cite{Lerche:1987sg,Lerche:1987qk} or more abstractly
through the elliptic genus \cite{Witten:1986bf}.

Because of its topological nature, the $B\wedge X_8$ term in (\ref{eq:apm3})
provides a useful handle on the study of higher derivative corrections in
dimensionally reduced IIA theories.  While there are often technical
difficulties involved for practical calculations, abstractly once the
corrections arising from $B\wedge X_8$ are pinned down, many of the remain
terms may be obtained by supersymmetry.  As an example of this, consider the
lift of IIA to eleven-dimensional supergravity.  In this case, $B\wedge
X_8$ lifts to $C_3\wedge X_8$, so that \cite{Duff:1995wd}
\begin{equation}
S_{11}=\fft1{2\kappa_{11}^2}\int\left[R*1-\ft12F_{4}\wedge *F_{4}
-\ft16C_{3}\wedge F_{4}\wedge F_{4}
+(4\pi\kappa_{11}^2)^{2/3}C_{3}\wedge X_{8}+\cdots\right],
\label{eq:d=11}
\end{equation}
where
\begin{equation}
X_{8}=\fft1{(2\pi)^4}\left(-\fft1{768}(\tr R^2)^2+\fft1{192}\tr R^4\right).
\label{eq:x8def}
\end{equation}
Compactifying to five dimensions on a Calabi-Yau three-fold gives
$\mathcal N=2$ supergravity coupled to $n_v$ vector multiplets and $n_h$
hypermultiplets with $n_v=h_{(1,1)}-1$ and $n_h=h_{(2,1)}+1$
\cite{Cadavid:1995bk}.

Focusing only on the vector multiplets, the compactification of
(\ref{eq:d=11}) on CY$_3$ proceeds by expanding the K\"ahler form $J$ in
a basis of $(1,1)$ forms $\omega_I$ on CY$_3$.  The Chern-Simons term then
reduces in a straightforward manner
\begin{equation}
\int_{\mathcal M_{11}} C_3\wedge F_4\wedge F_4 = \int_{\mathcal M_5}
c_{IJK}A^I\wedge F^J\wedge F^K,
\end{equation}
where $c_{IJK}$ are the triple intersection numbers.  Similarly, the
gravitational Chern-Simons term reduces as
\cite{Ferrara:1996hh,Antoniadis:1997eg}
\begin{equation}
\int_{\mathcal M_{11}} C_3\wedge X_8 = -\int_{\mathcal M_5}
\fft{c_{2I}}{24}A^I\wedge\tr R^2,
\end{equation}
where $c_{2I}$ arises from the expansion of the second Chern class
\begin{equation}
c_{2I} = \fft1{16(2\pi)^2}\int_{CY_3}\omega_I\wedge\tr R^2.
\end{equation}
The power of supersymmetry then enables us to deduce the entire
five-dimensional $\mathcal N=2$ action for the vector multiplets in terms
of the topological data $c_{IJK}$ and $c_{2I}$.  In particular, at the
$R^2$ level, the supersymmetric completion of $A^I\wedge\tr R^2$
was obtained in \cite{Hanaki:2006pj} using superconformal tensor calculus
and an off-shell formalism
\cite{Kugo:2000hn,Bergshoeff:2001hc,Fujita:2001kv,Bergshoeff:2004kh}.
The resulting bosonic action has the form \cite{Hanaki:2006pj}
\begin{eqnarray}
S_5&=&\fft1{2\kappa_5^2}\int\Bigl[
R*1-\ft12\mathcal N_{IJ}dM^I\wedge*dM^J-\ft12G_{IJ}F^I\wedge*F^J
-\ft16c_{IJK}A^I\wedge F^J\wedge F^K\nonumber\\
&&\kern4em-\fft{c_{2I}}{24}(\ft14A^I\wedge\tr R^2
-\ft18M^IC_{\mu\nu\rho\sigma}^2*1+\cdots)\Bigr].
\label{eq:5dimr2}
\end{eqnarray}
The addition of these $R^2$ terms lead to corrections to the entropy of
five-dimensional $\mathcal N=2$ black holes
\cite{Castro:2007sd,Castro:2007hc,Castro:2007ci}.  Furthermore, as the
$R^2$ terms are related to the five-brane anomaly, many of these entropy
results are in fact exact \cite{Harvey:1998bx,Kraus:2005vz,Kraus:2005zm}.

\subsection{$R^2$ corrections and AdS/CFT}

The supersymmetry analysis of \cite{Hanaki:2006pj} suggests that
any $R^2$ correction to five-dimensional $\mathcal N=2$ supergravity has
the form (\ref{eq:5dimr2}), with a precise relation between the
coefficient of $C_{\mu\nu\rho\sigma}^2$ and the gravitational Chern-Simons
term $A^I\wedge\tr R^2$.  Truncating to the pure supergravity sector and
integrating out the auxiliary fields of the off-shell theory, the
effective four-derivative action has the form \cite{Cremonini:2008tw}
\begin{eqnarray}
S_5&=&\fft1{2\kappa_5^2}\int\Bigl[
R*1-\fft32F\wedge*F+\fft{12}{L^2}*1+
\left(1-4\fft\alpha{L^2}\right)A\wedge F\wedge F\nonumber\\
&&\kern4em-\alpha\Bigl(\fft14A\wedge\tr R^2
-\fft18C_{\mu\nu\rho\sigma}^2*1+\cdots\Bigr)\Bigr].
\label{eq:5dimeff}
\end{eqnarray}
At this level, the theory is completely determined by two parameters: $L$,
the AdS radius and $\alpha$, the coefficient of the four-derivative correction
terms.  Note that we have chosen a non-canonical normalization for the
graviphoton which however is natural in the context of IIB supergravity
reduced on a Sasaki-Einstein manifold.

As highlighted above, for Calabi-Yau compactifications of eleven-dimensional
supergravity, $\alpha$ is given by the second Chern class of CY$_3$.
However, in an AdS/CFT setup, $\alpha$ also has a direct relation to the
central charges of the dual gauge theory.  This is perhaps best seen through
the holographic Weyl anomaly \cite{Henningson:1998gx}, where the
$\alpha C_{\mu\nu\rho\sigma}^2$ term in (\ref{eq:5dimeff}) shifts the leading
supergravity result
\cite{Nojiri:1999mh,Blau:1999vz,Imbimbo:1999bj,Fukuma:2001uf}.
Anomaly matching then yields the AdS/CFT connection
\cite{Buchel:2008vz,Cremonini:2008tw}
\begin{equation}
\fft{L^3}{\kappa_5^2}=\fft{a}{\pi^2},\qquad \fft\alpha{L^2}=\fft{c-a}{a},
\label{eq:acmap}
\end{equation}
where $L$ is the AdS radius, and where $a$ and $c$ are the central charges
of the dual $\mathcal N=1$ gauge theory.

Large $N$ theories with an AdS dual have leading behavior $a=c\sim N^2/4$
\cite{Henningson:1998gx}.  However, from (\ref{eq:acmap}) we see that
$1/N$ and further subleading corrections will show up in the dual
gravity theory as $R^2$ corrections parametrized by $\alpha$.  These
$R^2$ corrections have received much recent attention in computations of
the shear viscosity and consequences for the conjectured KSS bound
$\eta/s\ge1/4\pi$ for the ratio of the shear viscosity to the entropy density
of the dual gauge theory plasma \cite{Kovtun:2003wp,Kovtun:2004de}.
In particular, at linearized order, the ratio takes the form
\cite{Brigante:2007nu,Brigante:2008gz,Kats:2007mq,Buchel:2008vz,Cremonini:2008tw}
\begin{equation}
\fft\eta{s}=\fft1{4\pi}\left(1-\fft\alpha{L^2}+\cdots\right)
=\fft1{4\pi}\left(1-\fft{c-a}a+\cdots\right),
\end{equation}
so that theories with $c>a$ will violate the KSS bound.

While many examples of super-Yang Mills theories are known with $c\ne a$,
we are mainly interested in theories admitting a dual string description.
Several explicit examples have been constructed with, {\it e.g.}, seven-branes
and orientifolds, where $c-a\sim\mathcal O(N)$
\cite{Fayyazuddin:1998fb,Aharony:1998xz,Aharony:1999rz,Blau:1999vz,Aharony:2007dj}.  From the stringy point of view, the correction $\alpha$ in
(\ref{eq:5dimeff}) arises from the effective theory of the branes at the
singularities, and can be viewed as an open string effect that gives rise
to a $1/N$ correction to the leading $N^2$ behavior of the central charges.

In this paper, we wish to examine the closed string and hence $\mathcal O(1)$
corrections to $c-a$ by appropriate reduction of the higher derivative
terms in the bulk effective action.  Focusing on Sasaki-Einstein
compactifications of the IIB string, we immediately run into a puzzle.
Namely, as discussed above, the $\alpha$ correction term in
(\ref{eq:5dimeff}) is easily related to the reduction of the $C_3\wedge X_8$
term in eleven-dimensional supergravity or the corresponding $B_2\wedge X_8$
term in IIA theory.  However, it is well known that such a $B_2\wedge X_8$
term is absent in the IIB case, as the $(p,q)$ IIB fivebrane is non-chiral.
Thus, in contrast with (\ref{eq:apm3}), the $\alpha'^3$ corrections in IIB
have the schematic form
\begin{equation}
S_{\mathrm{IIB}}[\alpha'^3]=\fft{\alpha'^3}{2\kappa_{10}^2}\int d^{10}x
\sqrt{-g}\left[e^{-2\phi}(t_8t_8+\ft18\epsilon_{10}\epsilon_{10})R^4
+(t_8t_8+\ft18\epsilon_{10}\epsilon_{10})R^4\right],
\end{equation}
where in fact the tree and loop terms combine with non-perturbative
corrections into a modular-invariant form in terms of the IIB axi-dilaton.
When reduced on SE$_5$, it is clear that the above action will not yield
a non-vanishing $A\wedge\tr R^2$ term of the form given in (\ref{eq:5dimeff}).
Hence this suggests that $\alpha=0$, and therefore that all gauge theories
dual to IIB string theory on AdS$_5\times\mathrm{SE}_5$ will have a vanishing
$1/N^2$ correction to the central charges (but will still generically have
$1/\lambda^{3/2}$ corrections arising from $R^4$ terms in the dual theory).

The above argument for the absence of $1/N^2$ corrections, however, fails
to fully take stringy effects into account.  In particular, as we demonstrate
below, the finite volume of SE$_5$ leads to a non-vanishing contribution to
$\alpha$ which is not present in uncompactified IIB theory.  One way to see
this is to note that any Sasaki-Einstein manifold admits a preferred U(1)
fibration over a four-dimensional K\"ahler-Einstein base $B$
\begin{equation}
ds^2(\mathrm{SE}_5)=ds^2(B)+(d\psi+\mathcal A)^2,
\label{eq:secanonical}
\end{equation}
where $d\mathcal A=2J$, with $J$ the K\"ahler form on $B$.  This isometry
circle then allows us to relate IIB theory on AdS$_5\times\mathrm{SE}_5$
to IIA theory on AdS$_5\times B\times S^1$ via T-duality
\cite{Duff:1997qz,Duff:1998us}.  This circle furthermore allows us to
reduce first from ten to nine dimensions, and then from nine to five.  In
nine dimensions, the $B_2\wedge X_8$ term in IIA theory reduces to
$A_1\wedge X_8$, where $A_\mu=B_{\mu9}$.  Under T-duality, we thus see that
the compactified IIB theory necessarily has a similar term, however this
time with $A_\mu=g_{\mu9}$.  Independent of the duality frame, this term
reduces to five dimensions on the base $B$ to give rise to a generically
non-vanishing $1/N^2$ correction parameterized by $\alpha$ in
(\ref{eq:5dimeff}).

In fact, working at finite circle radius, we demonstrate that there are a
large class of gravitational and mixed Chern-Simons terms of the form
$A\wedge X_8$ which arise in string theory.  While some of these have been
identified previously, the full story appears to be as yet incomplete.
We thus begin in Section~\ref{sec:stringamp} with a reexamination of such
terms which arise from the one-loop CP-odd sector of type II string theory,
paying particular attention to the requirements of T-duality invariance.
Following this, in Section~\ref{sec:actualcomp} we compute the $\mathcal O(1)$
corrections to $c-a$ arising from the closed string sector of IIB theory on
AdS$_5\times \mathrm{SE}_5$.  Since these one-loop CP-odd terms are generic
in string theory, we conclude in Section~\ref{reduction} with a discussion
of similar corrections in AdS$_4$ and AdS$_3$ theories.  We also comment on
T-duality invariance in the presence of higher derivative terms in an
Appendix.

\section{One-loop CP-odd terms in string theory}
\label{sec:stringamp}

Before considering the reduction to five dimensions, we review the origin of
the $B\wedge X_8$ term in ten dimensions.  The structure of this CP-odd term
can be obtained from an explicit one-loop five-point computation following the
procedure outlined in \cite{Lerche:1987sg,Lerche:1987qk}.  In the RNS formalism
for the type II string, this parity violating term arises as a sum of two
contributions, one from the odd-even and the other from the even-odd spin
structure sector.  Focusing on the odd-even sector, the one-loop amplitude
may be set up with one vertex operator in the $(-1,0)$ picture and the
remaining four in the $(0,0)$ picture
\begin{eqnarray}
V^{(-1,0)}(k_0,\zeta^{(0)})&=&\zeta^{(0)}_{\mu\nu}
\delta(\gamma)\psi^\mu(i\overline\partial X^\nu+\ft12\alpha'k_0\cdot
\bar\psi\bar\psi^\nu)e^{ik_0\cdot X},\nonumber\\
V^{(0,0)}(k_i,\zeta^{(i)})&=&\zeta^{(i)}_{\mu\nu}
(i\partial X^\mu+\ft12\alpha'k_i\cdot\psi\psi^\mu)
(i\overline\partial X^\nu+\ft12\alpha'k_i\cdot\bar\psi\bar\psi^\nu)
e^{ik_i\cdot X},
\label{eq:5verts}
\end{eqnarray}
along with a picture changing operator $\delta(\beta)\psi\cdot\partial X$
in the left-moving sector.  It is conventional to take the first vertex to
be the antisymmetric tensor $B_{\mu\nu}$ and the remaining four to be
gravitons.  However, it should be noted that the NSNS fields $h_{\mu\nu}$,
$B_{\mu\nu}$ and $\phi$ all involve the same vertex operators, with the
only difference being the nature of the polarization tensors $\zeta^{(i)}$.

The five-point function of the vertex operators (\ref{eq:5verts}) vanishes
unless all ten fermion zero modes are soaked up in the odd spin structure
sector.  This gives rise to an amplitude
\begin{eqnarray}
\mathcal A&=&i\zeta^{(0)}_{\mu_0\nu_0}
\epsilon^{\mu_0\nu_0\lambda_1\mu_1\lambda_2\mu_2\lambda_3\mu_3\lambda_4\mu_4}
k^1_{\lambda_1}\cdots k^4_{\lambda_4}\zeta^{(1)}_{\mu_1\nu_1}\cdots
\zeta^{(4)}_{\mu_4\nu_4}\times\nonumber\\
&&\qquad\int\fft{d^2\tau}{2\pi\tau_2}\fft1{4\tau_2}
\int\fft{d^2z_1}{2\tau_2}\cdots\int\fft{d^2z_4}{2\tau_2}
\sum_a\left\langle\prod_{i=1}^4
(i\overline\partial X^{\nu_i}+\ft12\alpha'k_i\cdot\bar\psi\bar\psi^{\nu_i})
e^{ik_i\cdot X}\right\rangle_{\!\!a},\quad
\label{eq:cpoddamp}
\end{eqnarray}
where the sum is over the three even spin structures.  Upon integration
of the vertex operators, this takes the form
\begin{equation}
\mathcal A=i\int\fft{d^2\tau}{2\pi\tau_2}\fft1{4\tau_2}A(\bar q).
\label{eq:ellipticg}
\end{equation}
This integrand $A(\bar q)$ is
only a function of $\bar q=e^{-2\pi i\bar\tau}$, and computes the elliptic
genus \cite{Witten:1986bf}.  Proceeding either by direct computation or
through the elliptic genus \cite{Vafa:1995fj}, we then see that the
above amplitude gives rise to the $B_2\wedge X_8$ term, as expected.

Before proceeding, we note that the extra factor of $1/4\tau_2$ in
(\ref{eq:ellipticg}) arises from the zero mode contraction
\begin{equation}
\langle\overline\partial X^\mu\partial X^\nu\rangle=
-\fft{\alpha'}{8\pi\tau_2}\eta^{\mu\nu},
\label{eq:nczero}
\end{equation}
which we will revisit below in the compact case.  This factor is
crucial in showing that the amplitude is a total worldsheet derivative,
so that only the $\tau_2\to\infty$ boundary term contributes when integrating
$\tau$ over the fundamental domain \cite{Lerche:1987qk,Vafa:1995fj}.

Furthermore, as mentioned above,
the closed string vertex operators (\ref{eq:5verts})
encode $h_{\mu\nu}$, $B_{\mu\nu}$ and $\phi$ through the polarization
tensors $\zeta^{(i)}$.  Thus, at this linear order, it is clear that the
$B_2\wedge X_8$ term incorporates not just $R^4$ but also the full set of
NSNS fields according to
\begin{equation}
B_2\wedge X_8(R)\quad\to\quad B_2\wedge X_8(\hat R),
\end{equation}
where
\begin{equation}
\hat R_{\mu\nu}{}^{\lambda\sigma}=R_{\mu\nu}{}^{\lambda\sigma}
+\nabla_{[\mu}H_{\nu]}{}^{\lambda\sigma}-2\nabla_{[\mu}
\delta_{\nu]}^\sigma\nabla^\lambda\phi.
\label{eq:Rtorsion}
\end{equation}
While this expression is linearized in $H_3$ and $\phi$, we expect it to have
a non-linear completion, so that $\hat R_{\mu\nu\rho\sigma}$ becomes the
curvature of the connection with torsion $\hat\omega=\omega+H$.  This
completion is also needed in order to obtain a T-duality invariant combination
of $g_{\mu9}$ and $B_{\mu9}$ under circle reduction.  We discuss this
point further in the Appendix.

The simple replacement of $R(\omega)$ by $\hat R(\hat\omega)$, however, cannot
be the entire story, as the type II string receives contributions from both
the odd-even and even-odd spin structures.  These contributions are
essentially identical except for the important fact that $B_{\mu\nu}$,
being antisymmetric, has opposite worldsheet parity from $h_{\mu\nu}$
and $\phi$.  Noting that the IIA and IIB amplitudes differ by a relative
sign in the flip between odd-even and even-odd spin structures (because
of the differing GSO projection), we finally obtain the result for the
CP-odd sector
\begin{equation}
S_{\rm IIA}[\alpha'^3]=\fft{(2\pi)^6\alpha'^3}{2\kappa_{10}^2}
\int\bigl[B_2\wedge X_8(\hat R)\bigr]_{\mbox{odd in }B_2},
\label{eq:iiacorr}
\end{equation}
and
\begin{equation}
S_{\rm IIB}[\alpha'^3]=\fft{(2\pi)^6\alpha'^3}{2\kappa_{10}^2}
\int\bigl[B_2\wedge X_8(\hat R)\bigr]_{\mbox{even in }B_2}.
\label{eq:iibcorr}
\end{equation}
In particular, the $B_2\wedge X_8(R)$ term is projected out in the IIB case
by worldsheet parity.  However, terms of the form $B_2\wedge R(\nabla H)^3$
and $B_2\wedge R^3(\nabla H)$ survive.

\subsection{Reduction to $D=9$ and T-duality}

In the above, we have demonstrated the existence of a one-loop CP-odd
correction to IIB supergravity given by (\ref{eq:iibcorr}).  However,
this term by itself has no effect on the holographic $c-a$ computation,
as $H_3$ vanishes in the AdS$_5\times\mathrm{SE}_5$ background.  Instead, new
terms will show up when IIB string theory is compactified on a circle
to nine dimensions.  After all, from a IIA point of view, the familiar
$B_2\wedge X_8(\hat R)$ term in (\ref{eq:iiacorr}) may be reduced on a
circle in the $x^9$ direction to give $A_1\wedge X_8+B_2\wedge X_7$ where
$A_\mu=B_{\mu9}$ and $X_7$ is the circle reduction of $X_8$.  Focusing on the
first term, we note that T-dualizing to a IIB frame yields $A_1\wedge X_8$
where now $A_\mu=g_{\mu9}$.  Thus T-duality guarantees that IIB theory
on a circle necessarily includes a one-loop $A_1\wedge X_8$ term.  Of course,
this IIB CP-odd term does not lift to ten dimensions, as it would
schematically lift to $g_2\wedge X_8$ (where $g_2=\fft12
g_{\mu\nu}dx^\mu\wedge dx^\nu$), which however vanishes because of
the symmetry of the metric.

To directly see what is happening in nine dimensions, it is instructive to
return
to the five-point one-loop amplitude.  Assuming a circle of radius $R$ and
the first vertex in (\ref{eq:5verts}) to have a leg on the circle (so that
$\zeta_{\mu\nu}^{(0)}\to \zeta_{\mu9}^{(0)}$), we end up with a zero mode
contraction
\begin{equation}
\langle\overline\partial X^9\partial X^9\rangle=-
\left(\fft{\alpha'}2\right)^2\langle p_Lp_R\rangle.
\end{equation}
Here
\begin{equation}
p_L=\fft{n}R+\fft{wR}{\alpha'},\qquad
p_R=\fft{n}R-\fft{wR}{\alpha'},
\end{equation}
where $n$ and $w$ correspond to momentum and winding on the circle, and
the expectation is with respect to the partition function
\begin{equation}
Z=\tr q^{(\alpha'/4)p_L^2}\bar q^{(\alpha'/4)p_R^2}.
\label{eq:partf}
\end{equation}
This replaces the contraction (\ref{eq:nczero})
in the non-compact case that was used to obtain the additional factor of
$1/\tau_2$ in (\ref{eq:ellipticg}).

To see the effect of this circle compactification, we first consider the
large radius limit, $R\to\infty$.  In this case, only the zero winding sector
contributes, and we obtain
\begin{equation}
\langle\overline\partial X^9\partial X^9\rangle_{R\to\infty}
=-\fft{\alpha'}{8\pi\tau_2},
\label{eq:dbarx9dx9}
\end{equation}
up to exponentially suppressed corrections.  As expected, this directly
reduces to the non-compact zero-mode contraction (\ref{eq:nczero}).
On the other hand, in the small radius limit, $R\to0$, only the zero
momentum sector contributes, and we have instead
\begin{equation}
\langle\overline\partial X^9\partial X^9\rangle_{R\to0}
=\fft{\alpha'}{8\pi\tau_2}.
\end{equation}
The difference in sign is apparent from the opposite sign of the winding term
in $p_R$.

Combining the odd-even and even-odd spin structure sectors, we see that the
IIA CP-odd amplitude is proportional to the factor
\begin{equation}
B_{\mu9}(\langle\overline\partial X^9\partial X^9\rangle
+\langle\overline\partial X^\mu\partial X^\mu\rangle)
\sim\begin{cases}
B_{\mu9}&R\to\infty,\cr
0&R\to0,
\end{cases}
\label{eq:iiapv}
\end{equation}
while the IIB amplitude has the opposite behavior
\begin{equation}
g_{\mu9}(\langle\overline\partial X^9\partial X^9\rangle
-\langle\overline\partial X^\mu\partial X^\mu\rangle)
\sim\begin{cases}
0&R\to\infty,\cr
g_{\mu9}&R\to0.
\end{cases}
\label{eq:iibpv}
\end{equation}
Since T-duality relates large and small radius compactifications of IIA
and IIB theory, this result explicitly demonstrates the T-duality
covariance of $B_2\wedge X_8$ in nine dimensions with $B_2$ on the circle.
In fact, the extension of T-duality to the full reduction of $B_2\wedge X_8$
necessitates the use of the curvature with torsion (\ref{eq:Rtorsion})
and the fact that $B_2\wedge X_8(\hat R)$ is a top form, so that T-duality
will always flip between even and odd terms in $B_2$ in (\ref{eq:iiacorr})
and (\ref{eq:iibcorr}).

While the expressions (\ref{eq:iiapv}) and (\ref{eq:iibpv}) are appropriate
in the large and small radii limits, we are of course interested in
corrections arising at a finite radius.  With the supergravity limit
in mind, we work with $R$ finite and larger than $\sqrt{\alpha'}$.  As
in the $R\to\infty$ limit, only the zero winding sector contributes.  However,
approximating the momentum sum by an integral is only valid for $\tau_2\lesssim
R^2/\alpha'$.  For $\tau_2\gtrsim R^2/\alpha'$, the zero mode contraction
(\ref{eq:dbarx9dx9}) becomes exponentially suppressed.  Hence the radius
provides a natural cutoff
\begin{equation}
\langle\overline\partial X^9\partial X^9\rangle_{R^2\gg\alpha'}
\sim\begin{cases}\displaystyle-\fft{\alpha'}{8\pi\tau_2}&\tau_2\lesssim
R^2/\alpha',\\
0&\mbox{otherwise}.
\end{cases}
\end{equation}
The implication of this is that when the zero mode contraction takes place
on the $x^9$ circle the integral over the fundamental domain in
(\ref{eq:ellipticg}) is cut off at $\tau_2\sim R^2/\alpha'$.  So long as
$R^2\gg\alpha'$, the boundary contribution to the amplitude is still evaluated
at large $\tau_2$ and is hence dominated by the $\bar q^0$ term in the
elliptic genus $A(\bar q)$.  Integrating $\int d\tau_2/\tau_2^2$ up to
a cutoff of $R^2/\alpha'$ then demonstrates that the amplitude $\mathcal A$
picks up a finite radius correction factor of $1-\alpha'/R^2$ compared
to the non-compact result.

In addition, it is important to note that the discrete momentum sum for
the partition function on a circle will affect the amplitude even if the
zero mode contraction is in a non-compact dimension, as in (\ref{eq:nczero}).
In particular,
the bosonic zero mode contribution in (\ref{eq:partf}) takes the form
\begin{equation}
Z_{R^2\gg\alpha'}\sim\begin{cases}\displaystyle
\fft1{\sqrt{4\pi^2\alpha'\tau_2}}&\tau_2\lesssim R^2/\alpha',\\
\displaystyle\fft1{\sqrt{4\pi^2R^2}}&\mbox{otherwise}.
\end{cases}
\end{equation}
Since this contribution no longer falls off as $1/\sqrt{\tau_2}$ as
$\tau_2\to\infty$, the integral of the elliptic genus $A(\bar q)$ is
enhanced by a factor of $1+\alpha'/R^2$ whenever the zero mode contraction
$\langle\overline\partial X^\mu\partial X^\nu\rangle$ is over a non-compact
dimension.

Combining these two finite radius corrections, we may now refine the
above expressions (\ref{eq:iiapv}) and (\ref{eq:iibpv}) for the CP-odd
amplitudes.  In the large radius limit, the IIA amplitude is proportional to
\begin{equation}
(\langle\overline\partial X^9\partial X^9\rangle
+\langle\overline\partial X^\mu\partial X^\mu\rangle)A_1\wedge X_8
\sim\fft12\left[\left(1-\fft{\alpha'}{R^2}\right)
+\left(1+\fft{\alpha'}{R^2}\right)\right]A_1\wedge X_8=A_1\wedge X_8,
\end{equation}
where $A_\mu=B_{\mu9}$, while the IIB amplitude is proportional to 
\begin{equation}
(\langle\overline\partial X^9\partial X^9\rangle
-\langle\overline\partial X^\mu\partial X^\mu\rangle)A_1\wedge X_8
\sim\fft12\left[\left(1-\fft{\alpha'}{R^2}\right)
-\left(1+\fft{\alpha'}{R^2}\right)\right]A_1\wedge X_8=-\fft{\alpha'}{R^2}
A_1\wedge X_8,
\end{equation}
where $A_\mu=g_{\mu9}$.  What this indicates is that the $\alpha'^3$
corrections to the effective supergravity actions contain the following
CP-odd terms in nine dimensions%
\footnote{In principle, additional lower derivative terms such as
$A_1 \wedge \tr R^2 \wedge F^2$ could show up at finite circle radius. These
terms, however, arise from non-zero momentum or winding sectors (depending
on the T-duality frame), and hence are exponentially suppressed in the
compactification radius.}
\begin{eqnarray}
S_{\mathrm{IIA}}[\alpha'^3]&=&\fft{2\pi R_{\mathrm{IIA}}}{2\kappa_{10}^2}
(2\pi)^6\alpha'^3\int A_1\wedge [X_8(\hat R)]_{\mbox{even in }B_2},\nonumber\\
S_{\mathrm{IIB}}[\alpha'^3]&=&\fft{2\pi R_{\mathrm{IIB}}}{2\kappa_{10}^2}
\fft{(2\pi)^6\alpha'^4}{R_{\mathrm{IIB}}^2}\int A_1\wedge
[X_8(\hat R)]_{\mbox{even in }B_2},
\label{eq:ii9dim}
\end{eqnarray}
where of course the one-form potentials correspond to $B_{\mu9}$ and $g_{\mu9}$
for IIA and IIB, respectively.  Here we have explicitly written out the
nine-dimensional Newton's constant, and furthermore these expressions are
valid in
the corresponding large radii limits.  (For completeness, we note that there
are other CP-odd term as well, such as those associated with $B_2\wedge X_7$.
However, they vanish for the AdS$_5\times\mathrm{SE}_5$ reduction and hence
will not contribute to the $c-a$ computation.)

It is perhaps worth mentioning that the nine-dimensional CP-odd terms in
(\ref{eq:ii9dim}) are related by na\"ive T-duality where
$R_{\rm IIB}=\alpha'/R_{\rm IIA}$.  Thus we could have immediately written down
the $A_1\wedge X_8$ term for IIB theory compactified on a circle based
on the existence of the corresponding well-known term in IIA theory.
However, T-duality takes a large radius IIB theory into a corresponding
small radius IIA limit, in which case the IIA supergravity reduction is
not necessarily to be trusted so that a full string calculation is warranted.
Nevertheless, it is reassuring to see that the string amplitude calculation
and the supergravity reduction are in perfect agreement.

\section{Actual computation for AdS$_5\times \mathrm{SE}_5$}
\label{sec:actualcomp}

We now proceed to examine the $\mathcal O(1)$ contribution to $c-a$ for
$\mathcal N=1$
theories dual to IIB theory on AdS$_5\times \mathrm{SE}_5$.  Our strategy is
to take advantage of the fact that any Sasaki-Einstein metric may be given
in terms of a U(1) fibration over a K\"ahler-Einstein base $B$.  This allows
us to proceed in two steps: first reduce to nine dimensions on $S^1$, and then
further reduce down to five dimensions on the base $B$.  By taking the
intermediate step of working in nine dimensions, we may then straightforwardly
evaluate the nine-dimensional CP-odd term (\ref{eq:ii9dim}) to obtain
the $\mathcal O(1)$ contribution to $c-a$.

At the two-derivative level, the full non-linear reduction of the bosonic
sector of IIB theory on SE$_5$ was carried out in \cite{Buchel:2006gb}.
The dimensionally reduced fields $(g_{\mu\nu},A_\mu)$ comprise the bosonic
components of the $\mathcal N=2$ supergraviton multiplet in five
dimensions, and are related to the ten-dimensional fields according to
\begin{eqnarray}
ds_{10}^2&=&g_{\mu\nu}dx^\mu dx^\nu+L^2\left[ds^2(B)+(d\psi+\mathcal A
+L^{-1}A_\mu dx^\mu)^2\right],\nonumber\\
F_{5}&=&(1+*_{10})G_{5},\qquad
G_{5}=\fft4L\epsilon_{5}-L^2J\wedge *_5F_{2},\qquad F_2=dA_1.
\label{eq:blans}
\end{eqnarray}
Here we have written the Sasaki-Einstein metric as a U(1) bundle over $B$
\begin{equation}
ds^2(\mathrm{SE}_5)=ds^2(B)+(d\psi+\mathcal A)^2,\qquad d\mathcal A=2J.
\end{equation}
The resulting five-dimensional action is that of gauged $\mathcal N=2$
supergravity
\begin{equation}
S_5=\fft1{2\kappa_5^2}\int\left[
R*1+\fft{12}{L^2}*1-\fft32F_{2}\wedge*F_{2}+A_{1}\wedge F_{2}
\wedge F_{2}\right],
\label{eq:2dfive}
\end{equation}
where
\begin{equation}
\fft1{2\kappa_5^2}=\fft{L^5\mathrm{vol}(\mathrm{SE}_5)}{2g_s^2\kappa_{10}^2}.
\label{eq:5dnewton}
\end{equation}
Here $\mathrm{vol}(\mathrm{SE}_5)$ is the dimensionless volume of SE$_5$
and $2\kappa_{10}^2=(2\pi)^7\alpha'^4$.

The ansatz (\ref{eq:blans}) corresponds to the reduction of IIB theory on
a circle of constant radius $L$.  From a nine-dimensional point of view,
the fields are
\begin{eqnarray}
ds_9^2&=&g_{\mu\nu}dx^\mu dx^\nu+L^2ds^2(B),\nonumber\\
\tilde F_{2}&=&2LJ+F_{2},\nonumber\\
F_{4}&=&4L^3\epsilon_{4}(B)-L^2J\wedge F_{2},
\label{eq:nine}
\end{eqnarray}
where $\tilde F_2$ is the field strength of the Kaluza-Klein gauge field
$g_{\mu9}$.  This allows us to directly compute the one-loop CP-odd term
(\ref{eq:ii9dim}) which arises after circle compactification of IIB
theory.  Since $H_3$ vanishes for this background, we may take the
familiar $X_8$ given in (\ref{eq:x8def}).  Noting that the irreducible
$\tr R^4$ term does not contribute for the direct product nine-dimensional
metric (\ref{eq:nine}), we obtain
\begin{equation}
S_5[\alpha'^3]=-\left[\fft1{384}\fft{2\pi R}{2\kappa_{10}^2}
\fft{(2\pi)^2\alpha'^4}{R^2}\int_B\tr R^2\right]\int A_1\wedge\tr R^2.
\end{equation}
Comparing this expression with the general four-derivative action
(\ref{eq:5dimeff}), and using the five-dimensional Newton's constant relation
(\ref{eq:5dnewton}) allows us to extract the effective four-derivative
coefficient
\begin{equation}
\alpha=\fft{g_s^2}{96}\fft{2\pi R}{L^5\mathrm{vol}(\mathrm{SE}_5)}
\fft{(2\pi)^2\alpha'^4}{R^2}\int_B\tr R^2
\end{equation}
We now use the AdS/CFT relation $4\pi g_s N = L^4/\alpha'^2$ (which is
appropriate for the dual quiver gauge theories arising from a stack of
D3-branes at the tip of the cone over SE$_5$) and the holographic
anomaly matching relations (\ref{eq:acmap}) to write
\begin{equation}
a=\fft{N^2}4\fft{\mathrm{vol}(\mathrm{SE}_5)}{\pi^3},\qquad
c-a=\fft1{96}\fft{2\pi}{\mathrm{vol}(S^1)}\int_B\fft1{8\pi^2}\tr R^2,
\label{eq:aandc}
\end{equation}
where $\mathrm{vol}(S^1)$ is the dimensionless volume of the U(1) circle.
The $a$ anomaly expression is familiar%
\footnote{Note that we do not focus here on a possible overall $\mathcal O(1)$
shift $N^2\to N^2-1$ that is expected to show up in the expression for
$a$ and that has been computed through quantum corrections arising from
the Kaluza-Klein tower \cite{Bilal:1999ph,Mansfield:2002pa}.},
while the $c-a$ difference
picks up a calculable contribution from the closed string sector.

Before proceeding, it is important to keep in mind that $\tr R^2$ in the
expression for $c-a$ is composed out of the pullback of the ten-dimensional
curvature onto the base $B$.  In particular, since the compactification
manifold is a fibered space, the curvature of the U(1) bundle $d\mathcal A=2J$
will contribute as well the curvature of the base $B$.  In other words, the
ten-dimensional $\tr R^2$ will reduce to the nine-dimensional
$\tr\tilde R^2$ plus terms involving the Kaluza-Klein field $g_{\mu9}$
through its field-strength $\tilde F_2$.  If it were not for the latter
terms, then we would simply obtain
\begin{equation}
c-a=\fft1{96}\fft{2\pi}{\mathrm{vol}(S^1)}\int_B\fft1{8\pi^2}\tr \tilde R^2
=-\fft1{96}\fft{2\pi}{\mathrm{vol}(S^1)}\int_Bp_1
=-\fft{\sigma(B)}{32}\fft{2\pi}{\mathrm{vol}(S^1)},
\end{equation}
where $\sigma(B)=\int_Bp_1/3$ is the signature of the base $B$.  This,
however, is not the complete story, as the Kaluza-Klein gauge field is
non-trivial on the base as well.

\subsection{The Kaluza-Klein reduction of $\tr R^2$}

In order to evaluate the contribution of the Kaluza-Klein gauge field to
$\tr R^2$, we take the explicit circle reduction
\begin{equation}
ds_{10}^2=e^\alpha e^\alpha+e^{2\varphi}(d\psi+\tilde A)^2.
\label{eq:kkred}
\end{equation}
The resulting spin connections are
\begin{eqnarray}
\omega^{\alpha\beta}&=&\tilde\omega^{\alpha\beta}
-\ft12e^\varphi\tilde F^{\alpha\beta}
e^9,\nonumber\\
\omega^{\alpha9}&=&-\ft12e^\varphi\tilde F^\alpha{}_\beta e^\beta
-\partial^\alpha\varphi e^9,
\end{eqnarray}
where $\tilde\omega^{\alpha\beta}$ is the nine-dimensional spin connection
computed from the nine-dimensional metric $ds_9^2=e^\alpha e^\alpha$.
The curvature two-forms are then
\begin{eqnarray}
R^{\alpha\beta}&=&[\tilde R^{\alpha\beta}-\ft14e^{2\varphi}
(\tilde F^{\alpha\beta}\tilde F_{\gamma\delta}+\tilde F^\alpha{}_\gamma
\tilde F^\beta{}_\delta)e^\gamma e^\delta]\nonumber\\
&&\qquad-\ft12e^\varphi[\nabla_\gamma\tilde F^{\alpha\beta}
+2\tilde F^{\alpha\beta}\partial_\gamma\varphi+\tilde F^\alpha{}_\gamma
\partial^\beta\varphi-\tilde F^\beta{}_\gamma\partial^\alpha
\varphi]e^\gamma e^9,\nonumber\\
R^{\alpha9}&=&-\ft12e^\varphi[\nabla_\beta\tilde F^\alpha{}_\gamma
+\tilde F^\alpha{}_\gamma\partial_\beta\varphi+\tilde F^\beta{}_\gamma
\partial_\alpha\varphi]e^\beta e^\gamma\nonumber\\
&&\qquad+[\ft14e^{2\varphi}\tilde F^{\alpha\gamma}\tilde F_{\beta\gamma}
-\nabla^\alpha\nabla_\beta\varphi-\partial^\alpha\varphi\partial_\beta
\varphi]e^\beta e^9.
\end{eqnarray}

As indicated in (\ref{eq:nine}), for the U(1) fibered SE$_5$, the
nine-dimensional graviphoton field strength is a sum of two terms: the
K\"ahler form of $B$ and the five-dimensional graviphoton.  Since we are
primarily interested in extracting the coefficient of $A\wedge\tr R^2$ in
five dimensions, we ignore the graviphoton contribution.  In this case,
$\tilde F_2=2J$ is covariantly constant on $B$.  (Note that we have
explicitly scaled out the AdS radius factor $L$.)  Furthermore, the U(1)
circle has constant radius, so we set $\varphi=0$.  The curvature two-forms
then simplify as
\begin{equation}
R^{AB}=\begin{pmatrix}
\tilde R^{\alpha\beta}&0&0\\
0&\tilde R^{ab}-(J^{ab}J_{cd}+J^a{}_c J^b{}_d)e^c e^d&e^ae^9\\
0&-e^ae^9&0
\end{pmatrix}.
\end{equation}
Here we have further split the tangent space indices as
$\alpha,\beta=0,\ldots,4$ in five-dimensions, $a,b=5,\ldots,8$ on the
base $B$ and $9$ for the U(1) fiber.

A simple computation now demonstrates that
\begin{equation}
\tr R^2=[\tr\tilde R^2]_{\mathrm{AdS}_5}
+[\tr\tilde R^2+4\tilde R^{ab}J_{ab}J+2\tilde R^{ab}J_{ac}J_{bd}e^ce^d
-24J\wedge J]_{B}.
\end{equation}
Since $B$ is a K\"ahler manifold, we may use the identities
$\tilde R^{ab}J_{ac}J_{bd}=\tilde R_{cd}$ and $\tilde R^{ab}J_{ab}=2\rho$
(where $\rho$ is the
Ricci form) to simplify the second term above.  In this case we have
\begin{equation}
\tr R^2=[\tr\tilde R^2]_{\mathrm{AdS}_5}
+[\tr\tilde R^2+8\rho\wedge J-24J\wedge J]_{B}.
\end{equation}
Since we have scaled out the radius $L$, what remains is
a `unit radius' Sasaki-Einstein five-fold, where the Ricci curvature is
given by $[R_{ab}]_{\mathrm{SE}_5}=4\delta_{ab}$.  This five-dimensional
Einstein condition then requires the curvature of the K\"ahler-Einstein base
to satisfy $[\tilde R_{ab}]_B=6\delta_{ab}$, so that the Ricci-form is given by
$\rho=6J$.  This gives the final expression
\begin{equation}
\tr R^2=[\tr\tilde R^2]_{\mathrm{AdS}_5}+[\tr\tilde R^2+24J\wedge J]_{B}.
\label{eq:trr2red}
\end{equation}
Note that $J\wedge J$ is twice the volume form on the base.

We now return to the expression (\ref{eq:aandc}) for $c-a$ and substitute
in (\ref{eq:trr2red}) for the reduction of $\tr R^2$ on the base $B$.
The result is
\begin{eqnarray}
c-a&=&\fft1{96}\fft{2\pi}{\mathrm{vol}(S^1)}\int_B\fft1{8\pi^2}
\left(\tr\tilde R^2+24J\wedge J\right)\nonumber\\
&=&\fft1{32}\fft{2\pi}{\mathrm{vol}(S^1)}
\left(-\sigma(B)+\fft{2\,\mathrm{vol}(B)}{\pi^2}\right),
\label{eq:cminusa}
\end{eqnarray}
where $\mathrm{vol}(B)$ is the dimensionless volume of the base $B$.

\subsection{Reduction on $S^5$}

Given the above result, we now present a few examples where $c-a$ may be
directly computed.  We start with  IIB theory on AdS$_5\times S^5$, which
yields the familiar duality to $\mathcal N=4$ super-Yang-Mills theory.  In
this case, the
five-sphere may be written as $U(1)$ bundled over $CP_2$.  The signature
of $CP_2$ is $\sigma(CP_2)=1$, while its volume is $\pi^2/2$.  As a result,
the two terms cancel in (\ref{eq:cminusa}), and we verify that $c=a$ in
$\mathcal N=4$ super-Yang-Mills.  Alternatively, we may compute the
Riemann curvature from the Fubini-Study metric and directly show that
$\tr\tilde R^2=-24J^2$.  This demonstrates that it is not just the integrated
expression in (\ref{eq:cminusa}), but also the integrand itself that
vanishes everywhere on $CP_2$.

The result that $c-a$ remains unshifted by string loop contributions is of
course consistent with the expectation that while decoupling of the center
of mass U(1) takes U($N$) to SU($N$), this shift affects both $a$ and $c$
identically.  Thus $c$ remains identified with $a$, with $c=a=(N^2-1)/4$.
Note that, since the integrand of (\ref{eq:cminusa}) is trivial, this
string loop correction will continue to vanish for orbifolds of $S^5$,
such as $S^5/\mathbb Z_3$.  Since this model is dual to $\mathcal N=1$,
$\mathrm{SU}(N)^3$ gauge theory, it suggests that the $\mathcal O(1)$
contribution to $c-a$ ought to be $c-a=3/16$, corresponding to three
decoupled $\mathcal N=1$ vectors from the U(1) factors.  Although the
string loop calculation does not give such a shift, this could be
accounted for by a second contribution to $c-a$ arising from quantum
corrections from the states in the Kaluza-Klein tower
\cite{Bilal:1999ph,Mansfield:2002pa}.  In particular, the Kaluza-Klein
modes that need to be considered are those arising from the massless
fields in ten dimensions compactified on $S^5/\mathbb Z_3$, and it is
precisely these modes that have not been captured by the string loop
calculation.  While the contribution of complete $\mathcal N=8$
supergravity multiplets would not shift $c-a$, the orbifolding breaks
this to $\mathcal N=2$, in which case their contributions would no longer
be expected to vanish.  It of course remains to be seen whether a
direct computation along the lines of \cite{Bilal:1999ph,Mansfield:2002pa}
will reproduce the predicted value of $c-a=3/16$.

\subsection{Reduction on $T^{1,1}$}

The next simplest case to consider is IIB theory on AdS$_5\times T^{1,1}$
\cite{Klebanov:1998hh}.  Since the base of $T^{1,1}$ is $S^2\times S^2$,
we simply have $\tr\tilde R^2=0$.  (The vanishing of the signature can
also be understood from the existence of both a self-dual and an
anti-self dual harmonic two-form.)  In this case we are left with
the volume term $\mathrm{vol}(B)=(4\pi/6)^2=4\pi^2/9$.  Substituting this
into (\ref{eq:cminusa}) then gives the $T^{1,1}$ result
\begin{equation}
c-a=\fft1{24},
\label{eq:t11c-a}
\end{equation}
where we also used the fact that the volume of the U(1) fiber is $4\pi/3$.
Curiously, this is the contribution to $c-a$ for a free
$\mathcal N=2$ hypermultiplet or negative that of a vector multiplet%
\footnote{Although we believe the sign of the correction to be correct,
keeping track of the sign conventions for the CP-odd terms is rather subtle.
In principle, a reliable means of fixing the sign may be through the
reduction of $R^4$ in the CP-even sector, as the sign of the Weyl-squared
contribution in (\ref{eq:5dimeff}) is unambiguous.}.

It would be interesting to see how this result may arise from the perspective
of the dual gauge theory.  Generalizing the idea of the $S^5/\mathbb Z_3$
orbifold, we focus on $\mathcal N=1$, SU($N$) quiver gauge theories.  Since
we expect $c-a$ to count the number of decoupled $\mathcal N=1$ vectors,
the natural prediction would be
\begin{equation}
c-a=\fft{\mbox{(\# of nodes in the quiver)}}{16}.
\label{eq:c-aguess}
\end{equation}
This gives $c-a=1/8$ for the conifold gauge theory, which however does not
agree with (\ref{eq:t11c-a}).  While it is possible that we have lost a
factor of three in the string loop computation, we instead suggest as above
that there is a second contribution to $c-a$ from the Kaluza-Klein tower
so that
\begin{equation}
c-a = \left.\fft1{24}\right|_{\mbox{string loop}}+\left.\fft1{12}
\right|_{\mbox{supergravity loop}}=\fft18,
\end{equation}
where the supergravity loop contribution arises from the Kaluza-Klein
tower in five dimensions.  While it would be interesting to perform such
a calculation, in practice the non-trivial Kaluza-Klein spectroscopy on
$T^{1,1}$ \cite{Ceresole:1999zs} would appear to make this a challenge.

\subsection{Reduction on $Y^{p,q}$}

We now turn to reductions on the Sasaki-Einstein manifold $Y^{p,q}$, which
are dual to a large family of $\mathcal N=1$ superconformal quiver gauge
theories \cite{Gauntlett:2004yd,Martelli:2004wu,Benvenuti:2004dy}.  On
the IIB supergravity side, the Sasaki-Einstein manifold $Y^{p,q}$ has
topology $S^2\times S^3$.  The metric in canonical form (\ref{eq:secanonical})
is given by
\begin{eqnarray}
ds^2&=&\fft{1-cy}6(d\theta^2+\sin^2\theta d\phi^2)+\fft{dy^2}{w(y)q(y)}
+\fft{w(y)q(y)}{36}(d\beta+c\cos\theta d\phi)^2\nonumber\\
&&\qquad+\fft19[d\psi'-\cos\theta d\phi+y(d\beta+c\cos\theta d\phi)]^2,
\end{eqnarray}
where
\begin{equation}
w(y)=\fft{2(a-y^2)}{1-cy},\qquad q(y)=\fft{a-3y^2+2cy^3}{a-y^2}.
\end{equation}
The parameters $a$ and $c$ are to be chosen to avoid conical singularities
at the poles $y_1\le y\le y_2$ where $y_1$ and $y_2$ are the two smallest
roots of the cubic $a-3y^2+2cy^3=0$.  Note that $c=0$ corresponds to
$T^{1,1}$, while if $c\ne0$ we can rescale the coordinates to set $c=1$.

In order to compute $c-a$ given by (\ref{eq:cminusa}), we take the natural
vielbein basis
\begin{eqnarray}
&&e_1=\sqrt{\fft{1-cy}6}d\theta,\kern3.2em
e_2=\sqrt{\fft{1-cy}6}\sin\theta d\phi,\nonumber\\
&&e_3=\fft1{\sqrt{w(y)q(y)}}dy,\qquad e_4=\fft{\sqrt{w(y)q(y)}}6(d\beta
+c\cos\theta d\phi).
\end{eqnarray}
For the U(1) fibration, we have
\begin{equation}
\mathcal A=-\fft13[\cos\theta d\phi-y(d\beta+c\cos\theta d\phi)],
\end{equation}
and it is easy to verify that $d\mathcal A=2J$ where
$J=e_1\wedge e_2+e_3\wedge e_4$.

Given the above metric, we may directly compute $\tr\tilde R^2$ on
the base
\begin{equation}
\tr\tilde R^2=24\left[\fft{(1-ac^2)^2}{(1-cy)^6}-1\right]J\wedge J,
\end{equation}
where
\begin{equation}
\fft12J\wedge J=\fft{1-cy}{36}\sin\theta
d\theta\wedge d\phi\wedge dy\wedge d\beta
\end{equation}
is the volume form on the base.  In order to proceed, we need to integrate
$\tr\tilde R^2+24J\wedge J$ on the base $B$.  However, a difficulty arises
in that for generic values of $p$ and $q$ the manifold $Y^{p,q}$ is
irregular.  This means that while the above expressions are valid locally
the base $B$ is ill defined as a base manifold.  At best, for appropriate
values of $p$ and $q$ the Sasaki-Einstein space is quasi-regular, and the
base is then an orbifold.

Although the base $B$ may be ill defined, the five-dimensional Sasaki-Einstein
manifold itself is smooth and free of curvature singularities.  Thus instead
of computing the signature and volume of $B$ separately, we directly
evaluate the five-dimensional quantity
\begin{equation}
\tr R^2=\tr\tilde R^2+24J\wedge J=24\fft{(1-ac^2)^2}{(1-cy)^6}J\wedge J.
\end{equation}
Integrating this over the entire SE$_5$ then gives
\begin{equation}
\int_{\mathrm{SE}_5}\tr R^2\wedge\ft13d\psi'
=\fft{64\pi^3}3\ell(y_2-y_1)(1+c(y_2+y_1)),
\end{equation}
where $\ell=P_1/p=P_2/q$ is related to the period of the U(1) fiber in the
notation of \cite{Gauntlett:2004yd}.  As a result, we find
\begin{equation}
c-a=\fft1{96}\fft{2\pi}{\mathrm{vol}(S^1)^2}\int_{\mathrm{SE}_5}
\fft1{8\pi^2}\tr R^2\wedge\ft13d\psi'
=\fft1{8}\left(\fft{2\pi/3}{\mathrm{vol}(S^1)}\right)^2
\ell(y_2-y_1)(1+c(y_2+y_1)).
\end{equation}
Relating $\ell$ and the roots $y_1$ and $y_2$ to the Chern numbers $p$ and
$q$ finally gives
\begin{equation}
c-a=\fft1{48}\left(\fft{2\pi/3}{\mathrm{vol}(S^1)}\right)^2
\fft{p(4p^2-9q^2)+(2p^2+3q^2)\sqrt{4p^2-3q^2}}{p^2(p^2-q^2)}.
\label{eq:c-aypq}
\end{equation}
For $p$ and $q$ chosen appropriately, the square root becomes rational,
and the base $B$ is an orbifold.  In this case, orbits of the U(1) fiber
close, and we may take $\mathrm{vol}(S^1)=2\pi/3$, corresponding to
$2\pi$ periodicity of the $\psi'$ circle.  However, for irregular $Y^{p,q}$
the orbits do not close, and this suggests that $\mathrm{vol}(S^1)$ should
be taken to be infinite, in which case $c-a$ would vanish.  This difference
in behavior for quasi-regular versus irregular Sasaki-Einstein manifolds
appears rather unusual, and merits further investigation.

Based on the dual quiver gauge theory with $2p$ gauge groups, we may expect
from (\ref{eq:c-aguess}) that $c-a=p/8$.  If this is the case, then the
additional contribution from the Kaluza-Klein tower would have to compensate
for the rather unwieldy function of $p$ and $q$ appearing in (\ref{eq:c-aypq}).

\section{Graviphoton backgrounds and lower dimensional AdS reductions}
\label{reduction}

While we have mainly focused on AdS$_5$ reductions of IIB supergravity,
similar features arise when examining AdS$_4$ reductions of eleven dimensional
supergravity on Sasaki-Einstein seven-folds given by a non-trivial circle
fibration over a six-dimensional K\"ahler-Einstein base.  In particular,
we demonstrate that the reduction of $C_3 \wedge X_8$ gives rise to
four-dimensional couplings of the $\mathcal N=2$ graviphoton $T$ with the
curvature tensor of the form
\begin{equation}
R_{\mu\nu\lambda\sigma}T^{\mu\nu}T^{\lambda\sigma},
\end{equation}
and further compute its coefficient.

So far, we had not been concerned with nontrivial ten-dimensional graviphoton
backgrounds, since they do not give rise to AdS$_5$ compactifications.  However,
for AdS$_4$ reductions of eleven-dimensional supergravity on SE$_7$, the
ten-dimensional graviphoton is important, and some field redefinitions may be
required in order to write down a consistent (one-loop) ten-dimensional action.
To see how this arises, we first look at the reduction of $C_3 \wedge X_8$ to
ten dimensions on a non-trivially fibered circle.

It is convenient to work on  a twelve-dimensional manifold $Y_{12}$ whose
boundary is the eleven-dimensional spacetime  $X_{11} = \partial Y_{11}$.
We are interested in the case where $X_{11}$ is a circle fibration over
ten-dimensional spacetime $M_{10}$:  $ U(1) {\rightarrow} X \stackrel{\pi}{\rightarrow} M$.  (In turn, $M$ is a boundary to an eleven-dimensional
manifold $Y_{11}$).  The isometry is generated by a vector field $v$ and the
dual global connection one form is denoted by $e$:
\begin{equation}
\imath_v e =1\, , \qquad d \, e = \pi^* T\, ,
\end{equation}
where $T$ is the graviphoton field strength.

Now consider the circle reduction:
\beq\label{c3x8}
\int_X C_3 \wedge X_8(TX) = \int_{Y_{12}} G_4 \wedge X_8  \longrightarrow
\int_{Y_{11}} \imath_v [ G_4 \wedge {\tilde X}_8]\,,
\eeq
where the tilde eight-form is a polynomial of ten (rather than
eleven)-dimensional curvatures and the graviphoton $T$. Every
eleven-dimensional quantity respects the isometry, {\it i.e.}~has a vanishing
Lie derivative with respect to the vector $v$: $\mathcal{L}_v (.) =
(d \imath_v + \imath_v \, d) (.) = 0$. This means in particular that closed
forms upon reduction yield closed forms of lesser rank.

For now let us ignore the sources and take $d G_4 = 0$. Then
$\mathcal{L}_v G= 0$ allows to write
\begin{equation}
G_4 = \pi^* F_4 + \pi^* H_3 \wedge e\, ,
\end{equation}
where $F_4$ and $H_3$ are ten-dimensional RR and NSNS fluxes respectively
($\imath_v G = - H_3$). The closure of $G$ leads to a pair of ten-dimensional
equations:
\begin{equation}
d F_4 - H_3 \wedge T = 0  \, , \qquad d H_3= 0 \, .
\end{equation}
Similarly, 
\begin{equation}
X_8 = \pi^* {\tilde X}_8 + \pi^* {\tilde X}_7 \wedge e \, ,
\end{equation}
where ${\tilde X}_7 = - \imath_v X_8$. Note that all  quantities with
tildes are polynomials in the (ten-dimensional) curvature $R$ and the
graviphoton $T$: ${\tilde X}_n =  {\tilde X}_n ({\hat R}^{11})
= {\tilde X}_n (R^{10}, T)$.  The closure of $X_8$ leads to:
\begin{equation}
d {\tilde X}_8 - {\tilde X}_7 \wedge T = 0  \, , \qquad d {\tilde X}_7= 0 \, .
\end{equation}
Moreover locally $X_8 = d X_7^{(0)}$. If not only  $X_8$ but also the
descendant $X_7^{(0)}$ respects the isometry,
{\it i.e.}~$\mathcal{L}_v  X_7^{(0)}$, it follows that 
$$
\imath_v X_8 = - d (\imath_v  X_7^{(0)}) \, .
$$
One can show  ${\tilde X}_7 = -\imath_v X_8$ is not only closed, but is also
exact: ${\tilde X}_7  = d {\tilde X}_6(R,T)$.
Note that the horizontal eight-form ${\tilde X}_8 - T \wedge {\tilde X}_6 $
is closed. We shall give explicit expressions for the polynomials with tildes
shortly.

For now, let us get back to the reduction of the one-loop term \eqref{c3x8}:
\bea\label{c3x8-red}
 \int_{Y_{12}} G_4 \wedge X & \longrightarrow & \int_{Y_{11}}  F_4 \wedge {\tilde X}_7 \ + H_3 \wedge {\tilde X}_8 \cr
&&= \int_{Y_{11}}  d (F_4 \wedge {\tilde X}_6)  - d F_4  \wedge {\tilde X}_6  + H_3 \wedge  {\tilde X}_8 \cr
&&= \int_{Y_{11}}  d (F_4 \wedge {\tilde X}_6) + H_3 \wedge [ {\tilde X}_8  -  T \wedge  {\tilde X}_6] \cr
&&= \int_{M_{10}} F_4 \wedge {\tilde X}_6 \ + B_2 \wedge [ {\tilde X}_8  - T \wedge {\tilde X}_6]\,.
\eea
Note that the RR four-form appearing in \eqref{c3x8-red} satisfies the
correct Bianchi identity, $d F_4 - H_3 \wedge T = 0$, while $B_2$ comes
wedged with a closed eight-form, hence ensuring that the one-loop term
(in the absence of fivebrane sources) is invariant under the NSNS gauge
transformation $\delta B_2 \rightarrow d \Lambda_1$.

While further reductions are not important for our purposes, clearly we
can extend this discussion to the case of multiple (commuting) isometries of
the eleven-dimensional background. Let us discuss the reduction to nine
dimensions. Starting from eleven dimensions, we should consider now a pair
of isometries generated by $v^1$ and $v^2$.  Here we concentrate on
$B_1 = \imath_{v^1} \imath_{v^2} \, C_3$ couplings. Even though the SL(2)
doublets are not important for our purposes, and (in IIA language) we shall
set the pair of graviphotons to zero, let us have a look at the complete set
of lower dimensional descendants of $C_3 \wedge X_8$. Due to isometries we
can write
\begin{equation}
G_4 = \pi^* F_4 + \pi^* F_3^i \wedge e^i +  \pi^* H_2  \wedge e^1 \wedge e^2\,,
\end{equation}
where $H_2 = d B_1$. One can check:
\beq\label{BI}
d F_4 - F_3^i \wedge T^i = 0  \, , \qquad d F_3^i - \epsilon^{ij} H_2 T^j = 0    \, , \qquad d H_2= 0 \, .
\eeq
Similarly for $X_8$:
\begin{equation}
X_8 = \pi^* {\tilde X}_8 + \pi^* {\tilde X}_7^i \wedge e^i +
\pi^* {\tilde X}_6 \wedge e^1 \wedge e^2 \, ,
\end{equation}
where now ${\tilde X}_n =  {\tilde X}_n ({\hat R}^{11})  = {\tilde X}_n (R^{9}, T^1, T^2)$. The closure of $X_8$ leads to:
\begin{equation}
d {\tilde X}_8 -   {\tilde X}_7^i \wedge T^i = 0  \, , \qquad d {\tilde X}_7^i - \epsilon^{ij} {\tilde X}_6 \wedge T^j  = 0  \, , \qquad d {\tilde X}_6 = 0 \,.
\end{equation}
Let us introduce quantities:
\begin{equation}
I_5 = d^{-1}  {\tilde X}_6  \, , \qquad I_6^i =  d^{-1} [ {\tilde X}_7^i - \epsilon^{ij} I_5 \wedge T^j]  = 0  \, , 
\end{equation}
and note that due to $\epsilon_{ij} T^i \wedge T^j = 0$
\begin{equation}
d [ {\tilde X}_8 - I_6^i \wedge T^i ]  = d {\tilde X}_8 -   {\tilde X}_7^i \wedge T^i  = 0  \, .
\end{equation}
The result of the reduction of $C_3 \wedge X_8$ is then
\beq \label{nine}
\int_{M_9}  B_1 \wedge[ {\tilde X}_8 - I_6^i \wedge T^i ]   - F_4 \wedge I_5 + \epsilon_{ij} \,  F_3^i \wedge I_6^i \, ,
\eeq
where $F_4$ and $F_3^i$ satisfy the Bianchi identities \eqref{BI}. Once
again, in absence of fivebranes, the action is invariant under $\delta B_1
\rightarrow d \Lambda_0$. 

From now on, we shall consider only ten-dimensional theories with a single
non-trivial graviphoton. A quick comment about anomalies is in order. Indeed
the term  $B_2 \wedge [ {\tilde X}_8  - T \wedge {\tilde X}_6]$ (just like
its eleven-dimensional ancestor) is not invariant under ten-dimensional
diffeomorphisms. For that matter, even the closed eight-form is not;
${\tilde X}_8  - T \wedge {\tilde X}_6$ is invariant only under the combined
action of ten-dimensional diffeomorphisms and graviphoton U(1)
transformations. In the presence of NS5-branes, $d H = \eta (W_6
\hookrightarrow M_{10})$, the variation will produce a complicated expression 
\beq
\label{gravifive}
d^{-1} \delta d^{-1} [ {\tilde X}_8  - T \wedge {\tilde X}_6]\,,
\eeq 
restricted to the fivebrane worldvolume.  Note that from the other side
$ {\tilde X}_8$ is simply a sum of the usual (closed) $X_8$ polynomial
and a part that depends on the graviphoton, ${\tilde X}_8 = X_8(TM_{10})
+ X_8(R,T)$.  Hence in a trivial graviphoton background, $T=0$, we recover
the usual ten-dimensional anomaly inflow%
\footnote{With an abuse of notation, $X(TM_D)$ refers to forms constructed
of polynomials in $D$-dimensional curvatures. Quantities with tilde
$\tilde X$ are polynomials in $R$ and $T$.}.
Of course, eleven-dimensional anomaly cancellation requires contributions
from three sources --- the fivebrane anomaly, the variations from the bulk
$C_3 \wedge X_8(TM_{11})$ and the (modified) Chern-Simons term
$C_3\wedge G_4 \wedge G_4$ \cite{Freed:1998tg}. The prediction of this argument is that the
reduction of the latter should yield counterparts to \eqref{gravifive}%
\footnote{A circumstantial argument in favor of this is given by recalling
that the contribution from $C_3\wedge G_4 \wedge G_4$  is important when the
normal bundle is not trivial. It is not hard to check that when the normal
bundle  $N$ is trivial and $M_{10} = W_6 \times N$, the lift to eleven
dimensions is provided by fibering the M-theory circle over $N$. In this
situation the graviphoton field strength $T$ is a horizontal form on $T$ and
hence pulls back to zero on $W_6$. In other words, when the  normal bundle is
trivial, the  coupling $B_2 \wedge [ ({\tilde X}_8 - X_8(TM_{10}))
- T \wedge {\tilde X}_6 ]$ is invariant.}.

\subsection{Higher derivative couplings in AdS$_4$}

Before turning to the reduction, it is instructive now to look at explicit
expressions. Since the single trace part of $X_8$ does not contribute to the
reductions, it is sufficient to look at $X_4 = p_1(TM_{11})$ and
${\tilde X}_4$ and ${\tilde X}_3 = d {\tilde X}_2$ arising from the reduction:
\bea
8\pi^2{\tilde X}_4 &=& R^{ab} \wedge R^{ba} - ( R^{ab} T^{ba}) \wedge T
-\fft12 R^{ab} \wedge T^b \wedge T^a +  \frac{1}{4} T^{ab} T^{ba}\,
T \wedge T\cr
&& + \fft14 T^{ab}\,  T^b \wedge T^a \wedge T
+ \frac{1}{2} \nabla^c T^a \wedge  \nabla^d T^a \wedge e^c \wedge e^d\,, \cr
8\pi^2 {\tilde X}_3 &=& \left(R^{ab}  - \frac{1}{2}T^{ab} \, T
-\fft14 T^a \wedge T^b \right) \wedge e^c \, \nabla^c T^{ba}
+\frac{1}{2}  e^c \wedge \nabla^c T^a  \wedge(T^{ad} \, T^d)\,,\quad
\eea
where $T = \frac{1}{2} T_{ab} \, e^a \wedge e^b$ and $T^a = T^a{}_b \, e^b$
and the covariant derivative $\nabla$ is taken with respect to the Levi-Civita
connection.  (All the curvatures here are ten-dimensional, and $a,b,c,\ldots$
are ten-dimensional tangent space indices.)  One can now compute
\beq
{\tilde X}_2 = d^{-1} {\tilde X}_3 = \frac{1}{8\pi^2} \left( R^{ab}
-\fft14T^{ab}T - \frac{1}{4} T^a \wedge T^b\right)T^{ba}\,,
\label{eq:x2def}
\eeq
and see that it is invariant.  Hence ${\tilde X}_3$ is cohomologically trivial.

While we have reduced $C_3\wedge X_8$ from eleven dimensions, as highlighted
in (\ref{eq:iiacorr}), the string loop amplitude necessarily involves the
curvature of the connection with torsion (\ref{eq:Rtorsion}).  In particular,
we need to make the replacement $R \Rightarrow {\hat R}$ inside $\tilde X$
in order to account for $H$ contributions to the NSNS part of the couplings.
However, these contributions are not important in the SE$_7$ reduction since
we backgrounds have vanishing $H$.

We are now ready to discuss the graviphoton couplings in AdS$_4$.  From
the original eleven-dimensional point of view, the Sasaki-Einstein reduction
takes the form \cite{Gauntlett:2007ma}
\begin{eqnarray}
ds_{11}^2&=&g_{\mu\nu}dx^\mu dx^\nu+L^2[ds^2(B)+(d\psi+\mathcal A+(2L)^{-1}
A_\mu dx^\mu)^2],\nonumber\\
G_4&=&\fft6L\epsilon_4-\fft{L^2}2J\wedge*_4F_2,\qquad F_2=dA_1.
\label{eq:se7ans}
\end{eqnarray}
The resulting four-dimensional action corresponds to minimal gauged
$\mathcal N=2$ supergravity with AdS radius $L/2$
\begin{equation}
S_4=\fft1{16\pi G_4}\int\left[R*1+\fft{24}{L^2}*1-\fft12F_2\wedge*F_2\right].
\label{eq:ads4lag}
\end{equation}
In ten dimensions, we may view this solution as a reduction of IIA
supergravity on a
six-dimensional K\"ahler-Einstein manifold $B$ with a K\"ahler form $J$.
The Sasaki-Einstein seven-fold is then obtained by taking a U(1) bundle
over $B$ with $d\mathcal A=2J$.  From this point of view, the four-dimensional
Newton's constant is given by
\begin{equation}
\fft1{16\pi G_4}=\fft{L^6\mathrm{vol}(B)}{2\kappa^2}.
\label{eq:4newt}
\end{equation}

We are of course interested in the one-loop correction to (\ref{eq:ads4lag}).
In the absence of the NSNS $B$ field, the
relevant term in \eqref{c3x8-red} is $\int_{M_{10}} F_4 \wedge {\tilde X}_6$.
Taking into account the constant factors in (\ref{eq:iiacorr}), we find
\begin{equation}
S_4[\alpha'^3]=\fft{g_s^2(2\pi)^6\alpha'^3}{2\kappa^2}
\int_{\mathrm{AdS}_4\times B}F_4\wedge\tilde X_6
=-\fft1{96}\fft{g_s^2(2\pi)^6\alpha'^3}{2\kappa^2}
\int_{\mathrm{AdS}_4\times B}F_4\wedge\tilde X_4\wedge\tilde X_2.
\end{equation}
Focusing on the $\mathcal N=2$ graviphoton, $F_2$, we pick out the
component $F_4=-(L^2/2)J\wedge*_4F_2$ from (\ref{eq:se7ans}), so that
\begin{eqnarray}
S_4[\alpha'^3]&=&\fft1{192}\fft{g_s^2(2\pi)^6L^2\alpha'^3}{2\kappa^2}
\int_{\mathrm{AdS}_4\times B}*_4F_2\wedge\tilde X_2\wedge J\wedge\tilde X_4
\nonumber\\
&=&\fft1{16\pi G_4}
\fft1{192}\fft{g_s^2(2\pi)^6\alpha'^3}{L^4\mathrm{vol}(B)}
\int_{\mathrm{AdS}}*_4F_2\wedge\tilde X_2\int_BJ\wedge\tilde X_4,
\label{eq:s4a3calc}
\end{eqnarray}
where we have also used (\ref{eq:4newt}).

Making use of (\ref{eq:x2def}), and taking the AdS$_4$ graviphoton to be
$T=F_2$ gives
\begin{equation}
*_4F_2\wedge\tilde X_2=\fft1{8\pi^2}\fft1{12}[R_{\mu\nu\lambda\sigma}
F^{\mu\nu}F^{\lambda\sigma}-\ft12F^4-\ft14(F^2)^2]*_41.
\end{equation}
As a result, the four-dimensional action (\ref{eq:ads4lag}) picks up a
correction at the four derivative level
\begin{equation}
S_4[\alpha'^3]=\fft1{16\pi G_4}\int\alpha L^2\left(R_{\mu\nu\lambda\sigma}
F^{\mu\nu}F^{\lambda\sigma}-\ft12F^4-\ft14(F^2)^2+\cdots\right)*1,
\label{ads4-red}
\end{equation}
where the ellipsis denotes terms that we have not focused on.  See \cite{Myers:2010pk}  for a recent discussion of these terms.

The coefficient $\alpha$ may be extracted from (\ref{eq:s4a3calc})
\begin{equation}
\alpha=\fft1{192}\fft{g_s^2(2\pi)^4\alpha'^3}{L^6\mathrm{vol}(B)}
\fft1{24}\int_BJ\wedge\tilde X_4
=\fft1{192}\fft{g_s^2(2\pi)^2\alpha'^3}{L^6\mathrm{vol}(B)}
\fft1{48}\int_BJ\wedge(\tr\tilde R^2+32J\wedge J),
\end{equation}
where we have made use of the six-dimensional version of (\ref{eq:trr2red}).
Finally, if the AdS$_4$ geometry arose from a stack of $N$ M2-branes probing
a $\mathbb C^4/\mathbb Z_k$ singularity, we may
use the relation $L^6=32\pi^2g_s^2kN\alpha'^3$ to write
\begin{equation}
\alpha=\fft\lambda{N^2}\fft{1}{9\cdot8192\,\mathrm{vol}(B)}
\int_BJ\wedge(\tr\tilde R^2+32J\wedge J),
\end{equation}
where $\lambda=N/k$ \cite{Aharony:2008ug}.
This integral vanishes on $CP_3$ (the base for $S^7$), but is generally
non-zero.

Note that in addition to \eqref{ads4-red}, there are four-derivative order
terms that are in general moduli dependent. Indeed, for $B = b^i \omega^i$
where $\omega^i \in H^2(B)$, the reduction yields 
\beq
\int_{\mathrm{AdS}_4\times B} b^i \omega^i \wedge \tr R^2 \wedge (\tr R^2
-8\pi^2 T \wedge {\tilde X}_2) =  \int_{\mathrm{AdS}_4}  \alpha^i \, b^i
\, \tr R^2 \, ,
\eeq
where $\alpha^i = \int_B  \omega^i \wedge (\tr R^2 +8J\wedge\rho
-32J\wedge J)=\int_B\omega^i\wedge(\tr\tilde R^2+64J\wedge J)$.
For $\mathcal N=8$ reductions, {\it i.e.}~where $B = CP_3$, the second
cohomology is one-dimensional and $\omega = J$. Moreover in this case
$b$ is constant, and the resulting $\int_{\mathrm{AdS}_4}\tr R^2$
correction is non-dynamical.

\subsection{Reducing to AdS$_3$}

We conclude with a brief discussion of reductions to AdS$_3$.  In this
case, it is convenient to pass via
six dimensions. The lowest order in derivatives one-loop contributions
are well-studied in the context of IIA/heterotic duality with 16
supercharges (see e.g. \cite{Vafa:1995fj, Duff:1995wd}), and can be
collected into a Chern-Simons like term
\begin{equation}
B \wedge \Bigl( \mathcal{F}^T L \mathcal{F} - \tr R^2 \Bigr) \, ,
\label{eq:bff}
\end{equation}
where $\mathcal{F}^T = (T, G_2, F^I_2)$ with $T$ as above denoting the
IIA RR one-form (graviphoton), while $G_2$ and $F^I_2$ descend from the
RR 3-form (see \cite{Duff:1995wd} for details). The intersection matrix
is given by $L = [\sigma^1 \oplus d_{IJ}]$, where $d_{IJ}$ is in turn the
intersection matrix of the internal space $K$ and $I,J = 1,...,h^2(K)$. For
theories with 16 supercharges $K$ is a $K3$ surface, and $d_{IJ}$ has
signature $(3,19)$. We shall mostly ignore these modes and concentrate on
$T$ and $G_2$, since these are terms that also survive the truncation to
theories with lower supersymmetry.

Our discussion makes it clear that there are two types of higher-derivative (but still one-loop) corrections to this coupling. Indeed in ten dimensions we have both new terms involving $\nabla H$ and $H^2$ and terms involving $T$. While along internal directions these vanish, they should appear in the  six-dimensional effective theory. After integration by parts, the Chern-Simons
term (\ref{eq:bff}) takes the form
\begin{equation}
H \wedge d^{-1} \Bigl(d_{IJ}  \, F^I  \wedge F^J + T \wedge G_2  + 8\pi^2 ({\tilde X}_4 - T \wedge {\tilde X}_2) + \cdots  \Bigr) \, .
\end{equation}
The complete coupling should have $O(4,20)$ symmetry and hence the modified
curvature terms should be written in terms of $\mathcal{F}$ and not simply
$T$.  However we do not try to impose this here, and just write down the
terms that involve the graviphoton (and the ellipsis stands for the rest).
The reduction to AdS$_3$ can now be readily performed, and yields a correction
term of the form
\begin{equation}
\int_{\mathrm{AdS}_3} d^{-1} \Bigl( d_{IJ}  \, F^I  \wedge F^J + T \wedge G_2  + 8\pi^2 ({\tilde X}_4 - T \wedge {\tilde X}_2) + \cdots  \Bigr) \, ,
\end{equation}
containing higher derivative terms in in addition to the expected gauge and
gravitational Chern-Simons terms.


\subsubsection*{Acknowledgments}

This work was supported in part by the US Department of Energy under grant
DE-FG02-95ER40899.  JTL wishes to acknowledge the hospitality of the LPTHE
Jussieu and IPhT CEA/Saclay; RM thanks MCTP for hospitality.  We wish to thank M. Petrini, P. Vanhove and B. Wecht
for stimulating discussions.


\appendix


\section{T-duality and higher-order terms in the effective action}

We shall present here a very brief (and incomplete) discussion of the action
of T-duality on the corrections to the effective action. Since the terms in
the effective action built solely out of curvature cannot be invariant under
T-duality, the corrections discussed in Section~\ref{sec:stringamp} can also
be seen as completions required to make higher curvature terms invariant. 

One way of introducing a correction with torsion is to consider the Courant
bracket---a generalization of the Lie bracket acting on vector fields, which
however acts on sections of the so called generalized tangent bundle $E$. The
latter locally is a product of tangent and cotangent bundles to the manifold
$M$ ($\mbox{dim}(M) = d$):
\begin{equation}
\label{eq:Edef}
   0 \longrightarrow T^*M \longrightarrow E 
      {\longrightarrow} TM \longrightarrow 0 . 
\end{equation}
Sections of $E$ are called generalized vectors. Locally they can be
written as $X=x+\xi$ where $x\in TM$ and $\xi\in T^*M$. In going from
one coordinate patch $U_\alpha$ to another $U_\beta$, we have to first
make the usual patching of vectors and one-forms, and then give a
further patching describing how $T^*M$ is fibered over $TM$ in
$E$. The choice of $B$ gives a canonical identification of $E$ with
$T \oplus T^*$, and the ordinary Courant bracket on $E$ gets identified with
the twisted Courant bracket on $T \oplus T^*$.

The generalized connection is defined by analogy to the ordinary connection
and is an operator
\begin{equation}
D : C^\infty(W) \to C^\infty(E\otimes W)  \, ,
\end{equation}
where $W$ is some vector bundle which carries a representation of
$O(d,d)$. We can now think of $D$ as $D=\partial+\Omega$, where the
ordinary derivative $\partial$ simply gives a term in the $T^*M$ part of
$E$ and nothing in the $TM$ part. Thus one defines the derivative
$D$, acting on a generalized vector $X$. The generalized connection is
now defined by the Courant bracket:
\begin{equation}
\label{eq:Courant}
   [ x + \xi, y + \eta ] 
     = [x,y]_{\rm Lie} + \mathcal{L}_x\eta - \mathcal{L}_y\xi
        - \frac{1}{2} d \left(\imath_x\eta - \imath_y\xi\right) ,
\end{equation}
where $[x,y]_{\rm Lie}$ is the usual Lie bracket between vectors and
$\mathcal{L}_x$ is the Lie derivative. 

We are interested in a case when the string background admits an isometry,
hence both the metric on  $M$ and $H$ are annihilated by the Lie derivative
of some vector $v$, $\mathcal{L}_{v} g = \mathcal{L}_{v}H = 0$. We shall use
the setup similar to that of Section~\ref{reduction}.  In particular we may
use $H = \pi^* H_3 + (\pi^* H_2) \wedge e$, where $\imath_v e = 1$ and
$d e = \pi^* F$.

We may also decompose the sections of $TM\oplus T^*M$ into horizontal and
vertical components, $x \longrightarrow x + f v  $ and $\rho \longrightarrow
\rho + \phi e$ correspondingly, and consider the Courant bracket
\begin{eqnarray}
\label{red-courant}
& [(x + f v ;   \rho + \phi e),   \,(y + g  v;   \lambda +  \omega e )]_{(H_3,H_2)} = \nonumber \\
& \qquad [(x;  \rho ),  \,(y; \lambda)]_{H_3} + 
 \Bigl(0 + ({\cal L}_{x} g - {\cal L}_{y} f) \, v ; \xi_{\mbox{base}}+
 ({\cal L}_{x} \omega - {\cal L}_{y} \phi) e   \Bigr)   \, ,
\end{eqnarray}
where the first term is the Courant bracket on the base of the circle
fibration and 
\begin{eqnarray}
\xi_{\mbox{base}} =&     \left (\begin{array}{cc }    \imath_{x} F  &  \imath_{x} H_2    \end{array}  \right )   \eta\      \left (\begin{array}{c} g   \\   \omega  \end{array} \right ) -  \left (\begin{array}{cc }    \imath_{y} F  &  \imath_{y} H_2    \end{array}  \right )   \eta\      \left (\begin{array}{c} f  \\   \phi  \end{array} \right ) \\ \nonumber
&+ 
 \left (\begin{array}{cc }    g  &  \omega   \end{array}  \right )   \eta\      \left (\begin{array}{c} d f  \\   d \phi  \end{array} \right )  - \frac{1}{2} d \Bigl(  \left (\begin{array}{cc }    g  &  \omega   \end{array}  \right )   \eta\      \left (\begin{array}{c}  f  \\    \phi  \end{array} \right )  \Bigr) \, ,
\end{eqnarray}
with $\eta =  \left (\begin{array}{cc }  0 & 1 \\ 1& 0  \end{array}  \right )$.
One can now readily check that an $O(1,1)$ transformation
\beq
\label{auto-dd}
\left (\begin{array}{c}  F\\ H_2  \end{array} \right )  \rightarrow  X  \cdot \left (\begin{array}{c}  F \\  H_2   \end{array} \right )
  \qquad  
   \mbox{and} \qquad    \left (\begin{array}{c}  f \\ \phi   \end{array} \right )  \rightarrow   X  \cdot \left (\begin{array}{c}  f \\  \phi  \end{array} \right )\,,  \quad \left (\begin{array}{c}  g \\ \omega   \end{array} \right )  \rightarrow   X  \cdot \left (\begin{array}{c}  g \\  \omega  \end{array} \right) 
\eeq
leaves $\xi_{\mbox{base}}$ invariant and is an automorphism of the bracket \eqref{red-courant}, provided that $X$ is an $O(1,1)$ matrix $X^T \eta X = \eta$. Hence the connection $\omega + H$ defined by \eqref{red-courant} is T-duality invariant. Finally, by flipping the sign of $H$, we may remark that the connection $\omega - H$ is T-duality anti-invariant.

This can be generalized for the case of multiple commuting isometries $v_I$,
provided $\imath_{v_I} \imath_{v_J} H = 0$ for any two vectors $v_I$ and $v_J$.
In general, for $n$ isometries, $O(n,n)$ transformations are not an
automorphism of the Courant bracket, and hence one cannot construct an
$O(n,n)$ invariant generalized (twisted) connection. More details on
generalized connection can be found in \cite{G-poisson, TdGCG, GMPW}.

One may also use the Courant bracket to define a curvature operator that
will be tensorial when restricted to integrable maximally isotropic
subbundles of $E$. For our purpose, it suffices to look at the curvature
${\hat R}$ written in \eqref{eq:Rtorsion} in the linearized approximation.
Once more, ${\hat R}_+ = {\hat R}(\omega + H)$ is T-duality invariant,
while ${\hat R}_- = {\hat R}(\omega - H)$ does not transform particularly
nicely under T-duality. However  writing locally $X_{4n} ({\hat R_\pm}) =
dX^{(0)}_{4n-1} (\omega \pm H)$, we recall that the latter contain only odd
powers of the connection, and hence  $X_8({\hat R}_+)$ and $X_8({\hat R}_-)$
are respectively even and odd under T-duality.

With all this in mind, the CP-odd corrections \eqref{eq:iiacorr},
\eqref{eq:iibcorr} and \eqref{eq:ii9dim} can be summarized as the
T-duality invariant combination
\beq
\label{coupl-T}
\left (\begin{array}{cc }  \gamma & \beta  \end{array}  \right )  \eta
\left (\begin{array}{c}  X_8({\hat R}_+) +  X_8({\hat R}_-)  \\
X_8({\hat R}_+) -  X_8({\hat R}_-)   \end{array} \right )  \, ,
\eeq
where the curvature expressions are constructed out of the original
ten-dimensional fields. From the IIA point of view, we have introduced
$\beta = B_{\mbox{inv}} + e\wedge \imath_v B$, with $ B_{\mbox{inv}}
= (1 -  e\wedge \imath_v ) B - \frac{1}{2} \imath_v g \wedge \imath_v B$
being the component of the $B$-field invariant under T-duality. We have also
introduced $\gamma = (\alpha'/R^2)  e \wedge \imath_v g  $ which vanishes in
the absence of isometries and is suppressed in the large radius IIA
ten-dimensional limit. Hence in ten dimensions the formula reproduces the known
CP-odd one-loop terms. Upon reduction on a circle, \eqref{coupl-T} correctly
reproduces the nine dimensional couplings.  Note that T-duality exchanges
the roles of $\imath_v B$ and $\imath_v g$, so that the former becomes
associated with $\gamma$ and the latter with $\beta$ from the IIB point of view.
Of course, this cannot be the complete story,
as it ought to be possible to express the CP-even corrections in a T-duality
invariant manner as well.  However, this looks somewhat more involved, as
the circle reduction of $\hat R_\pm$ appears rather unenlightening.


\end{document}